**The development of lower-atmosphere turbulence early in a solar flare**

N. L. S. Jeffrey[1*], L. Fletcher[1], N. Labrosse[1], and P. J. A. Simões[1]

[1]School of Physics & Astronomy, University of Glasgow, G12 8QQ, Glasgow, UK.

*Correspondence to: natasha.jeffrey@glasgow.ac.uk

**Abstract**
We present the first observational study of the onset and evolution of solar flare turbulence in the lower solar atmosphere on an unprecedented timescale of 1.7s, using the Interface Region Imaging Spectrograph observing plasma at a temperature of 80000 K. At this time resolution, non-thermal spectral line broadening, indicating turbulent velocity fluctuations, precedes the flare onset at this temperature, and is coincident with net blue-shifts. The broadening decreases as the flare brightens, then oscillates with a period of ~10s. These observations are consistent with turbulence in the lower solar atmosphere at the flare onset, heating that region as it dissipates. This challenges the current view of energy release and transport in the standard solar flare model, suggesting that turbulence partly heats the lower atmosphere.

**Introduction**
Solar flares are the dramatic result of magnetic energy release and dissipation in the Sun's atmosphere. This energy release and transfer is initiated by magnetic reconnection in the solar corona, but how it proceeds is still actively debated. During and following reconnection, random motions of the magnetized plasma - magnetohydrodynamic (MHD) plasma turbulence - may play a vital role in converting magnetic energy to the kinetic energy of accelerated particles and heating the flare corona to temperatures over 10 million Kelvin (MK) (1). In the standard solar flare model, the energy from this remote energy-release site in the corona is then transported by particles alone to heat the lower solar atmosphere. However, turbulent energy may also be dissipated and heat multiple regions throughout the flaring atmosphere (2), from the corona to the chromosphere (3).

The shape of an emitted spectral line can be used to understand the velocity distribution of the emitting particles. Early flare studies using full-Sun X-ray spectroscopy found that, during the early phase of some flares, spectral lines from >10 MK plasma exhibited line widths in excess of what is expected from random thermal motions alone (4,5), the so-called non-thermal line broadening. While line broadening can be caused by other mechanisms, such as opacity- or pressure-broadening, these should be negligible in optically thin solar flare plasma (6). The excess line broadening is thought to result from the superposed Doppler-shifted emission of turbulent fluid motions.

Spatially resolved observations with the Hinode (7) Extreme ultraviolet Imaging Spectrometer (EIS, 8) have greatly increased our knowledge of the role turbulence plays during a flare. Significant non-thermal broadening is found in solar active regions, even in the pre-flare stages (9). In flares, (10) found that non-thermal broadening increased with temperature and height in the coronal loop top source. Another study showed that the non-thermal broadening of one line, formed at 16 MK in the corona, peaked before the acceleration of electrons in the 'impulsive' phase of the flare (11). Further, the ratio of the turbulent kinetic energy to accelerated electron power gave a timescale for dissipating this turbulence of 1-10s, similar to predictions of MHD models (12). Such studies provide convincing observational evidence that MHD turbulence can act as a key energy transfer intermediary in the corona, heating plasma and converting magnetic energy into the kinetic energy of flare-accelerated electrons.

Large line broadening (~100 km/s) resulting from such turbulence is not unexpected in high-temperature regions close to coronal sites of flare energy release. However, do turbulent velocity fluctuations exist in the lower atmosphere, as well as the corona, during a flare? The Interface Region Imaging Spectrograph (IRIS, 13) permits the detailed study of flares in the chromosphere and the transition region (the thin atmospheric layer connecting the corona to the chromosphere). Both EIS and IRIS have observed non-thermal broadenings of over 50 km/s, in lines emitted in the heated lower flaring atmosphere, at temperatures ranging from 10,000 K to 10 MK (6,14). However, as they are often accompanied by "chromospheric evaporation" (15,16), which is a net up-flow of expanding lower-atmosphere plasma heated to well over 1 MK, disentangling turbulence and unresolved flows in this expansion can be difficult, particularly from data with low temporal resolution.

To investigate turbulent velocity fluctuations before and at the flare onset, prior to significant energy deposition and heating, we need to observe the rapid initial flare evolution at as high a temporal resolution as possible. Using IRIS data, we report the first high time-resolution spectroscopic study of early flare turbulence in the lower solar atmosphere. The observation captures this flare with an unprecedented 1.7 s time resolution and a high spatial resolution of 0.33" (~240 km).

**Results**

We observe a B-class flare (SOL2016-12-06T10:36:58) in active region 12615 (Fig. S1). Our analysis covers the flare from pre-flare to decay. Two bright flare ribbons, a signature of energy dissipation in the lower atmosphere, are observed in the extreme ultraviolet at 131 Å from the Solar Dynamics Observatory (SDO) Atmospheric Imaging Assembly (AIA) (17), during the impulsive phase. The Reuven Ramaty High Energy Solar Spectrometer Imager (RHESSI) (18) observes X-ray emission up to 9 keV. Integrating over a 4-minute period shows a 6-12 keV source located close to the western ribbon (Fig. 1).

The temporal evolution of one spectral line, Si IV 1402.77 Å (laboratory wavelength), formed at ~80000 K (see Materials and Methods, see Fig. S2), is examined. We study the period between 10:36:51 UT (+870 s from the IRIS start time of 10:22:22 UT) and 10:39:02 UT (+1000 s). During this time, the IRIS slit sits over the eastern flare ribbon, moving <1" in the east-west direction, so that it is observing essentially the same location at all times presented here. Having determined that Si IV is optically thin (see Materials and Methods), we analyze the line using single-Gaussian line fitting and a non-parametric moments analysis to determine the emitting plasma motions. We measure the line centroid positions, giving the bulk flow velocities v, relative to an inferred reference wavelength of (1402.775±0.020) Å, with an uncertainty corresponding to ± ~4 km/s (see Materials and Methods.). The width of the line is determined as a Gaussian Full-Width at Half-Maximum (FWHM).

Fig. 2 displays example spectral lines and their Gaussian fits at six times, while Fig. 3 shows the temporal evolution of the centroid positions (bulk flow velocities v), and widths (non-thermal motions $v_{nth}$), for a spatially integrated region covering 2" in the north-south direction, alongside the Si IV integrated intensity and RHESSI 6-12 keV X-ray light curve. The non-thermal width of Si IV begins to increase at 900 s, well before the Si IV intensity starts to rise. It peaks at 910 s, co-temporal with the start of the rise of Si IV intensity. Expressed as the non-thermal velocity, the line width increases from a pre-flare level of $v_{nth}$≈9 km/s to a peak of $v_{nth}$≈30 km/s at 910 s, with the line intensity peaking 20 s later, at 930 s. By 940 s, when the RHESSI 6-12 keV emission first peaks, $v_{nth}$ has returned to pre-flare levels.

During the peak phase, $v_{nth}$ displays periodic variations which can be fitted with a sinusoidal function, estimating period P≈11.5 s and amplitude A≈3.4 km/s. In the 30 s prior to the rapid rise in $v_{nth}$ at the start of the flare, the line wavelength moves systematically to the blue. From the onset of the rapid rise in $v_{nth}$ at 900 s, this blue-shift increases rapidly over the next 10 s to -18 km/s, stopping abruptly as $v_{nth}$ reaches its first peak. The blue-shift then decreases taking the line centroid back to its reference wavelength.

**Discussion**

Our interpretation is as follows. First, the evolution of the line intensity, non-thermal line width and line centroid position provides evidence that turbulent velocity fluctuations are indeed present in the lower solar atmosphere before the flare, contributing to plasma heating. The 30 km/s line broadening precedes the flare onset as indicated by its impulsive radiation signatures observed in X-rays and (extreme) ultraviolet (although we cannot rule out gentle heating between 900-920 s). The Si IV intensity only begins to increase, indicating lower atmospheric heating, as the broadening reaches its peak, consistent with the idea that the dissipation of turbulent energy over ~10 s contributes to heating of this region during the flare. The presence of turbulent signatures over ~60 s means that the driver of turbulence persists for longer. Si IV emission in flares is usually well-correlated with flare hard X-rays (HXR) [19], which provide the most direct signature of intense energy deposition by flare-accelerated electrons in the lower atmosphere. In this flare, however, there is no detectable high-energy HXR emission, though the 6-12 keV X-rays rise rapidly as the Si IV intensity peaks, possibly indicating low-energy non-thermal electrons.

Second, after its initial growth, the periodic variations of ~10 seconds in the line broadening are consistent with the growth and decay of a spectrum of interacting waves (turbulent velocity fluctuations) viewed over a spatially integrated region along the line of sight (see Fig. 4 and Materials and Methods). Due to instrumental constraints in previous missions, such rapid changes in broadening during a flare have not been observed previously (quasi-periodic oscillations have previously been observed in Si IV intensity and velocity [20], but on periods of minutes). The growth, variation and decay of the broadening coincide with movements in the Si IV centroid position, indicating small line-shifts, most notably a blue-shift of 18 km/s co-temporal with the initial rise of $v_{nth}$. Since they occur before the intensity has reached its peak and decrease as the flare intensity increases, these shifts are unlikely to be due to chromospheric evaporation. However, the observed patterns of $v_{nth}$ and $v$ can both be produced simultaneously by modeling the passage of turbulent velocity fluctuations in the region. The passage of a single fluctuation (or wave) moving through a region, with varying properties (e.g. amplitude, wavelength) cannot reproduce the observed temporal patterns in $v_{nth}$ and $v$, since this would produce a clear oscillation in the bulk velocity $v$ (Fig. 4), but a spectrum of interacting fluctuations can reproduce the observation results (Fig. 4, Materials and Methods). The spectral line Mg II 2796 Å, formed at lower temperatures, shows a similar bulk velocity pattern in the flare region, but with smaller blue-shifts of only 2-3 km/s (see Materials and Methods and Fig. S3), possibly due to decreasing fluctuations at lower altitudes. Modeling such fluctuations also suggests that the size of the region emitting Si IV along the line of sight is comparable to the mean wavelength of the fluctuations (Fig. 4, see Materials and Methods).

High time resolution measurements of the non-thermal and bulk velocities together provide detailed information about the nature of solar flare turbulence that can be compared with modeling, and emphasize the need for missions capable of sub-second spectroscopy. The results challenge the common view about the initiation of solar flares, and our understanding of energy release and transport in astrophysical plasmas. This first report of rapid oscillatory behavior in the line broadening of Si IV in the flaring lower-atmosphere is consistent with anisotropic plasma motions in the lower atmosphere becoming isotropic as the flare progresses (see Fig. 4, Materials and Methods) and strongly implies that the turbulent dissipation of magnetic energy might occur throughout the flare loop, far from the flare energy release sites in the corona, just before and during the flare onset.

## Materials and Methods

### The IRIS observation of Si IV 1402.77 Å

IRIS is a space-borne ultraviolet (UV) spectrograph that provides high-resolution spectroscopy in three wavelength ranges of 1332-1358 Å, 1389-1407 Å and 2783-2834 Å. It also provides slit-jaw imaging at four bands (C II 1330 Å, Si IV 1400 Å, Mg II k 2796 Å and Mg II wing 2830 Å). During the studied observation, the position of the IRIS slit only moves an angular distance of ~7" (~5,000 km), from solar east to west, in ~ 1 hour. The IRIS raster uses a rapid exposure time of 0.5 s, a slit movement time of ~1.7 s, and a field-of-view of 7" x 128". The pixel size in the north-south direction (Y) is 0.33" (~240 km). For the data analysis, we use prepped level-2 IRIS data and we removed any cosmic ray spikes. The data is then analyzed using the Interactive Data Language (IDL) and Solar SoftWare (SSW) routine iris_getwindata.pro. We study the strong line of Si IV 1402.77 Å (laboratory wavelength), which has no blends, and it is formed at the transition region temperature T [K] of logT=4.9 ~ 80000 K (peak formation temperature – see Fig. S2).

Between the flare times of ~10:37 UT and 10:40 UT, the IRIS slit only moves a small east-west distance of X<1", so we can study emission from approximately the same location at the start of the flare. We analyze the first three moments of the Si IV 1402.77 Å line distribution using: (1.) a non-parametric moments analysis that calculates the line distribution integrated intensity (I; zero moment), centroid ($\lambda_0$; first moment) and variance ($\sigma$; second moment), using:

$$I = \sum_{i=1}^{N} I_i, \quad \lambda_0 = \frac{\sum_{i=1}^{N} I_i \lambda_i}{I}, \quad \sigma = \frac{\sum_{i=1}^{N} I_i (\lambda_i - \lambda_0)^2}{I},$$

where $\lambda$ denotes the wavelength [Å], and (2.) single Gaussian line fitting,

$$I_i = A[0] + A[1] \exp\left(-\frac{(\lambda_i - A[2])^2}{2A[3]^2}\right)$$

where A[j] are the fitted parameters. Unlike other flare studies, where the determination of line shape was a key part of the analysis (21), here a detailed shape analysis is not required. We are only interested in an estimate of the random plasma motions along the line-of-sight (e.g., the second moment), and their temporal evolution. Hence, we approximate the entire line shape using a single Gaussian, even if the profile consists of multiple components. Since the number of line components is unknown, and the fitting of multiple Gaussians is not well constrained, it is sensible to estimate the velocity of all random turbulent motions together using a single Gaussian in this study. This is compared with the values determined from the moments analysis, but we find that both studies produce near identical results (Fig. 3, Fig. S4).

### Determining a reference wavelength for Si IV 1402.77 Å

For the determination of the line centroid velocity (or bulk velocity, v), we require a reference wavelength, that determines the average Si IV wavelength observed by IRIS in 'quiet' non-flaring regions during the observation. This is not necessarily the same as the Si IV 1402.77 Å laboratory wavelength provided by the CHIANTI line list (22,23), but the reference wavelength acts as a zero point for the calculation of the line centroid velocities in this observation. Often a reference wavelength for Si IV 1402.77 Å can be determined using photospheric lines with negligible flows, but none were suitable for determining a reference wavelength during this observation. Next, we attempted to determine a reference wavelength by obtaining the centroid position of Si IV 1402.77 Å from a region of 'quiet Sun'. However, we also found that the Si IV 1402.77 Å signal was very low and barely above the background noise in regions away from the flare, even with a large north-south Y binning of ~50". Finally, since IRIS observes the active region for a time of approximately 1 hour, we calculated the average centroid position of Si IV 1402.77 Å over all observation times excluding the time of the flare between 10:37 UT to 10:40 UT, in our studied region, close to the flare, using a Y binning of 10". From this, we determine a reference wavelength value of (1402.775±0.020) Å (Fig. 5), that is very close to the laboratory wavelength. All the Si IV line centroid velocities in Fig. 2, Fig. 3, Fig. 5, Fig. S3, and Fig. S4 are found relative to this reference wavelength. Converted to a velocity [km/s], the reference wavelength uncertainty is ±~4 km/s. We note that many studies find an average Si IV red-shift of 5-10 km/s in active regions (24), but we do not see any evidence of a red-shift here.

## The optical depth of Si IV 1402.77 Å

In certain environments, Si IV 1402.77 Å might suffer from radiative absorption effects. For example, self-absorption features have been observed in Si IV lines during transient brightenings (25). Hence, we test whether Si IV is likely to be optically thin for this study. If available, the Si IV 1393.76 Å /1402.77 Å doublet can be used to check the optical depth (26). Unfortunately, the Si IV 1393.76 Å line is not available in this observation. Therefore, we make an estimate of the optical depth τ using τ (λ)= $τ_0(λ_0)$ Φ(λ) (27) where $τ_0(λ_0)$ is the optical depth at the line center, $λ_0$ [Å] is the line centroid wavelength, and Φ(λ) is the absorption profile. Absorption effects are larger at the line center, and hence we only calculate the optical depth at the line center using $τ_0$~0.26 f <$n_e$>/$10^{10}$ [$cm^{-3}$], where f is a filling factor and $n_e$ is the ionized electron number density [$cm^{-3}$].

Next, we use the radiative hydrodynamics code RADYN (28,29) to model the possible flare atmospheric properties and to estimate the electron density $n_e$ where Si IV could be formed. To estimate the hydrodynamic response of the atmosphere, and its properties such as temperature and electron number density, RADYN requires an input energy flux [erg $s^{-1}$ $cm^{-2}$], that is usually estimated from the electron beam parameters determined from RHESSI X-ray analysis. We attempted to perform an analysis of the RHESSI spectral data for this flare. However, the low RHESSI sensitivity for this flare, the low count rate and the small detectable energy range, made the spectral analysis challenging. Nevertheless, sensible nonthermal electron parameters were estimated and the resulting RADYN inputs and outputs are shown in Fig. 6. Here we used model 84 from the [F-CHROMA flare model database](). Model 84 is chosen since it is based on a weak and soft nonthermal electron beam, that is appropriate for this flare. The RADYN results suggest that, at early times, Si IV is likely formed over a large range of electron number densities of log $n_e$ =9-12. Plugging these values into $τ_0$~0.26 f <$n_e$>/$10^{10}$ [$cm^{-3}$], provides an optical depth $τ_0$ in the range of 0.0007-20.0 (Fig. 6). We also use the filling factor (f) values of 0.1 and 1.0, and the line width (FWHM) between the thermal width of 0.05 Å and 0.2 Å, since $τ_0$ is inversely proportional to the FWHM. Fig. 6 shows that for log $n_e$ <10, Si IV is indeed optically thin (i.e. $τ_0$<1) for most of f and FWHM, and for most of log $n_e$ >10. Therefore, we cannot rule out weak absorption effects completely, but the analysis suggests that Si IV is optically thin for most of the possible flaring atmospheric conditions, and suitable for an analysis of turbulence in the lower atmosphere.

## The emission of Si IV 1402.77 Å in non-ideal conditions

*Non-equilibrium ionization:* Non-equilibrium ionization effects can cause peak formation temperatures to shift by log T=0.3 K in densities of $n_e$=$10^{10}$ $cm^{-3}$, but the shift should be negligible for larger densities (31), and line emission can be formed over a greater height range (32). Further, (33) studied C IV, an ion formed at a similar temperature to Si IV (log T=5.0) via modeling, and they suggest that temperature gradients and flow velocities can disrupt ionization equilibrium, and we might expect such conditions during a flare (e.g. chromospheric evaporation). However, this is a weak flare and we do not see any strong evidence for large flows before and during the flare, in the studied region.

*Multi-thermal plasma:* In the analysis, we assume that the average temperature of the region is close to the peak formation temperature of Si IV, and hence, we remove the thermal broadening associated with this temperature (~80000 K). However, Si IV, like other spectral lines, can be formed over a range of temperature. The contribution function (G(T); Fig. S2), shows that the emission of Si IV falls by one order of magnitude between the temperatures of 50000 K to 140000 K, and such temperatures would cause the thermal width to only vary between 5-9 km/s, which is negligible compared to the observed non-thermal line broadening of 30 km/s. So even if non-equilibrium ionization effects force Si IV to be formed at higher temperatures, this cannot account for the observed broadening (~30 km/s), and hence the formation of Si IV in a multi-thermal plasma does not change the result or the interpretation of the result.

## A comparison with the Mg II k-line (2796.35 Å)

In this observation, Si IV 1402.77 Å is the only optically thin line suitable for the analysis. However, we can compare the temporal evolution of the Si IV 1402.77 Å centroid position with the centroid positions of another available line, the Mg II k-line 2796.35 Å, formed at the cooler temperature T [K] of log T=4.5. We did not perform a detailed study of the Mg II line here, since the Mg II h and k resonance lines tend to have a double peak, due to effects of reabsorption. Hence, in most formation conditions Mg II is optically thick, and it can be formed over multiple layers in the low atmosphere, from the temperature minimum region to the upper chromosphere (34). Therefore, we can only estimate the line centroid positions (and hence relative bulk velocities) using a moments analysis. We note that the skewness (the third moment of the distribution) of the Mg II k line changes between values of (-0.1, +0.1) during flare times and it is possible that this could affect the interpretation of the line centroid position during the flare. In Fig. S3, we compare the temporal evolution and changes in centroid position for both Si IV and Mg II. The changes that we observe in the Si IV centroid position seem to be repeated for Mg II but to a much lesser extent. The magnitudes of the Si IV line velocities are larger with values between -18 km/s and 5 km/s (±3 km/s) compared to Mg II velocity range of -2 km/s to 3 km/s. We note that Si IV is difficult to analyze outside of the studied region, but Mg II is a stronger line in outside 'quiet' regions. Hence, we test whether the same changes in the Mg II centroid position are observed in different regions away from the flare, which could suggest an instrumental cause. We do not observe such centroid changes in the 'quiet' region (Fig. S3), and hence this suggests that the changes in centroid position, and therefore also width, are indeed physical in this observation.

## The Si IV 1402.77 Å centroid and width analysis

In the main text, we analyzed the centroid position and line broadening over a spatially integrated region made up of seven pixels, in the north-south direction, from Y=-114.9" to Y=-112.9". We also analyzed the seven pixels that make up this region individually. Compared to the spatially integrated case, we find that each of the seven individual pixels show a similar temporal evolution in their centroid positions and non-thermal line broadenings. In Fig. S4, we show the temporal evolution of centroid position and FWHM for two of the individual pixels: pixel 2 and pixel 4 located at Y=-114.2" and Y=-113.6", respectively. In all cases (Fig. 3 and Fig. S4), the line widths are determined from the second moment of the distribution and for the Gaussian fitting we determine the Gaussian Full Width at Half Maximum (FWHM) using 2 A[3] $(2\ln(2))^{1/2}$ or as $2\sigma (2\ln(2))^{1/2}$ from the moments analysis. The total line FWHM is given by FWHM=$(FWHM_{inst}^2 + 4\ln(2) (\lambda/c)^2 (v_{th}^2 + v_{nth}^2))^{(1/2)}$, where $FWHM_{inst}$ denotes the IRIS instrumental broadening [Å], c the speed of light [km/s], $v_{th}$ [km/s] the thermal Doppler velocity of the parent Si IV ions and $v_{nth}$ [km/s] the average non-thermal velocity of the plasma, along the line of sight. We determine the non-thermal line velocity by removing the IRIS instrumental broadening with a velocity of 3.9 km/s (13), and the Si IV thermal velocity of $v=(2k_BT/M)^{1/2}$=6.86 km/s, in quadrature, where $k_B$ is the Boltzmann constant [erg/K], and M is the mass of the Si ion [g]. The line centroid is converted to a Doppler velocity using $v=c(\lambda-\lambda_0)/\lambda_0$ where $\lambda_0$ is the centroid wavelength. The inferred centroid positions and FWHMs of Si IV in the integrated region of interest over all observation times are shown in Fig. 5.

**Modeling the observed temporal trends of $v_{nth}$ and v**

To try and reproduce all of IRIS observations (of bulk velocity and non-thermal velocity) together, we set up a simple model that represents the region emitting Si IV during the flare. We stress that we do not attempt to model the detailed transition region environment, or the line emission itself but *only* the plasma motions in this region, at some angle $>0°$ to the observer's line-of-sight. For this we set up a region of extent z, parallel to the guiding magnetic field (see Fig. 4).

We then set up a wave with wavelength $\lambda(t)$, amplitude $A(t)$ and wave speed v, and these properties vary with time, t, as the wave(s) passes through the region z, causing the material to fluctuate at 90° to z. We pass either a single wave or multiple waves with different properties. The transverse velocity of wave *i* takes the simple form of

$$\sum_{i=1}^{N} V_i(t,z) = \sum_{i}^{N} A_i(t) \cos\left(\frac{2\pi}{\lambda_i(t)}\right)\left(z - \frac{v}{i}t\right).$$

A sum of multiple waves above represents interacting waves (possibly interacting Alfvén waves) that can form plasma turbulence, and drive a cascade of wave energy to smaller and smaller wavelengths causing e.g. electron heating in the region.

The size of the region (z) emitting Si IV is unknown, so we define the region in terms of the average fluctuation wavelength $<\lambda>$, i.e. as $z/<\lambda>$.

The model uses a time resolution of 1.7s, matching the IRIS observations. At each time, we calculate the first moment and second moment of all transverse velocity fluctuations along z, which represent the bulk v and non-thermal velocities $v_{nth}$, respectively.

This simple modeling suggests several interesting results:

1. We can reproduce all the IRIS observations v and $v_{nth}$, if we input a spectrum of waves (~10 different modes). If the number of modes is too large, then we lose the detectable oscillations in $v_{nth}$, at this time resolution. This limits the number of interacting modes that can produce the observations.
2. The input of a single wave with varying properties cannot reproduce the observations. A single fluctuation will produce noticeable and resolved oscillations in v, which is not observed with IRIS.
3. We can only reproduce the observation if $z/<\lambda>$ is slightly less than 1 (~0.5). Situations where $z/<\lambda> >> 1$ seem to show large $v_{nth}$ but negligible v, while situations where $z/<\lambda> << 1$ seem to show very large v but negligible $v_{nth}$.

To match the observations, we require a spectrum of interacting waves, and this produces an initial transverse velocity field with greater anisotropy than a single passing wave, becoming more and more isotropic with time.

From the spectral line observations, at the start of the flare, the line profiles do show some anisotropy, but overall, they are approximately symmetrical (Fig. 2). Therefore, this suggests that the turbulence is unresolved by the IRIS observations over a single pixel distance of ~240 km.

However, we do not know the extent of the region emitting Si IV, along the line-of-sight. Again, the contribution function G(T) (Fig. S2) might suggest that Si IV could be formed over a temperature of 50000-140000 K, within one order of magnitude of its peak emission, and in the RADYN model atmosphere, this corresponds to a restricted height range of ≤100 km, at the start of the flare. Hence, if z is indeed of the order ≤100 km, then the above modeling suggests that $<\lambda> \leq 200$ km. However, the location of this temperature range is dynamic as a function of height throughout the simulations, but it is generally >1.5 Mm above the canonical photosphere "surface", or mid to top chromosphere and transition region. Therefore, our back of the envelope calculations should be viewed with caution, and we stress that a more detailed study, beyond the scope of this report, is required to understand the role of waves and turbulence in the lower atmosphere during the flare. Moreover, we cannot say whether such fluctuations are generated locally in the region, or if they are generated and transported from another location, such as the corona, during the flare energy release process.

**Acknowledgments**

**General**: We thank Peter J. Levens for obtaining the observation and Paul J. Wright for the use of his colorblind friendly color table.

We thank Eduard P. Kontar, Hugh S. Hudson and David Ireland for reading the manuscript and providing insightful science comments that greatly improved the text.

CHIANTI is a collaborative project involving George Mason University, the University of Michigan (USA) and the University of Cambridge (UK).

IRIS is a NASA small explorer mission developed and operated by LMSAL with mission operations executed at NASA Ames Research center and major contributions to downlink communications funded by ESA and the Norwegian Space Centre.

**Funding:** NLSJ, LF, NL gratefully acknowledge the financial support by STFC Consolidated Grants ST/L000741/1 and ST/P000533/1. PJAS acknowledges support from the University of Glasgow's Lord Kelvin Adam Smith Leadership Fellowship.

**Author contributions:** NLSJ performed data analysis, interpreted the results and produced the manuscript and figures. LF interpreted the results and helped to produce the manuscript. NL helped with the manuscript and interpretation of the results. PJAS performed data analysis, produced figures and helped to produce the manuscript.

**Competing interests:** There are no competing interests.

**Data and materials availability:** All data needed to evaluate the conclusions in the paper are present in the paper and/or the Supplementary Materials. Additional data related to this paper may be requested from the authors.

**Figures and Tables**

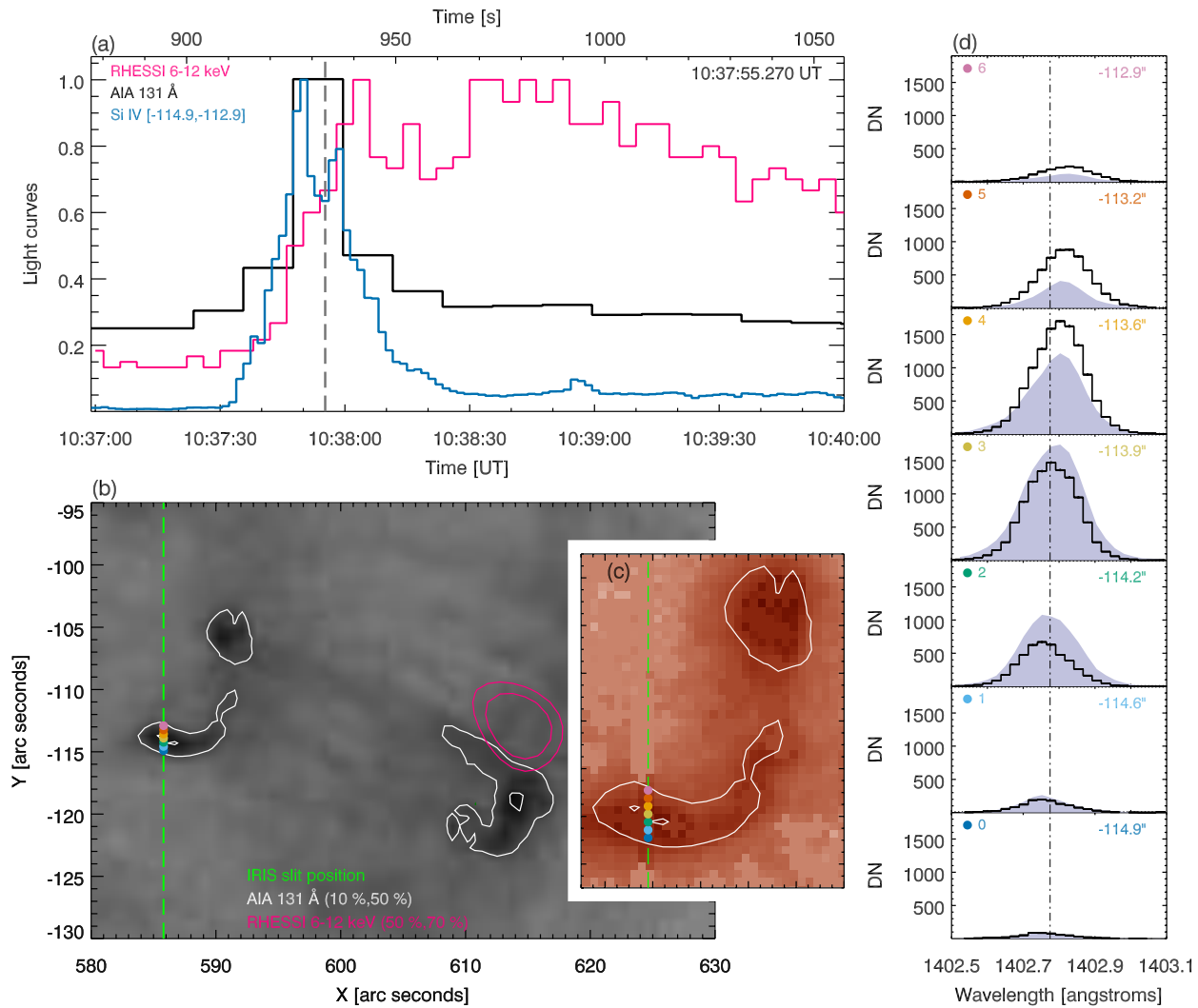

**Fig. 1.** The light curves, and the images and line spectra of flare SOL2016-12-06T10:36:58 at 10:37:55 UT. (a) The Si IV 1402.77 Å (blue), RHESSI 6-12 keV (pink) and AIA 131 Å (black) light curves. (b) AIA 131 Å background image with contours (white) showing the IRIS slit position (green) and RHESSI 6-12 keV contours (pink). (c) The eastern ribbon using an IRIS slit-jaw (SJI) 1400 Å image overlaid with AIA 131 Å contours. (d) The Si IV spectral lines at seven locations (colored dots in (b) and (c)); black line: current time and violet solid: previous time. The dashed-dot line shows the position of the inferred reference wavelength of (1402.775±0.020) Å. A movie showing all times is available (Movie 1).

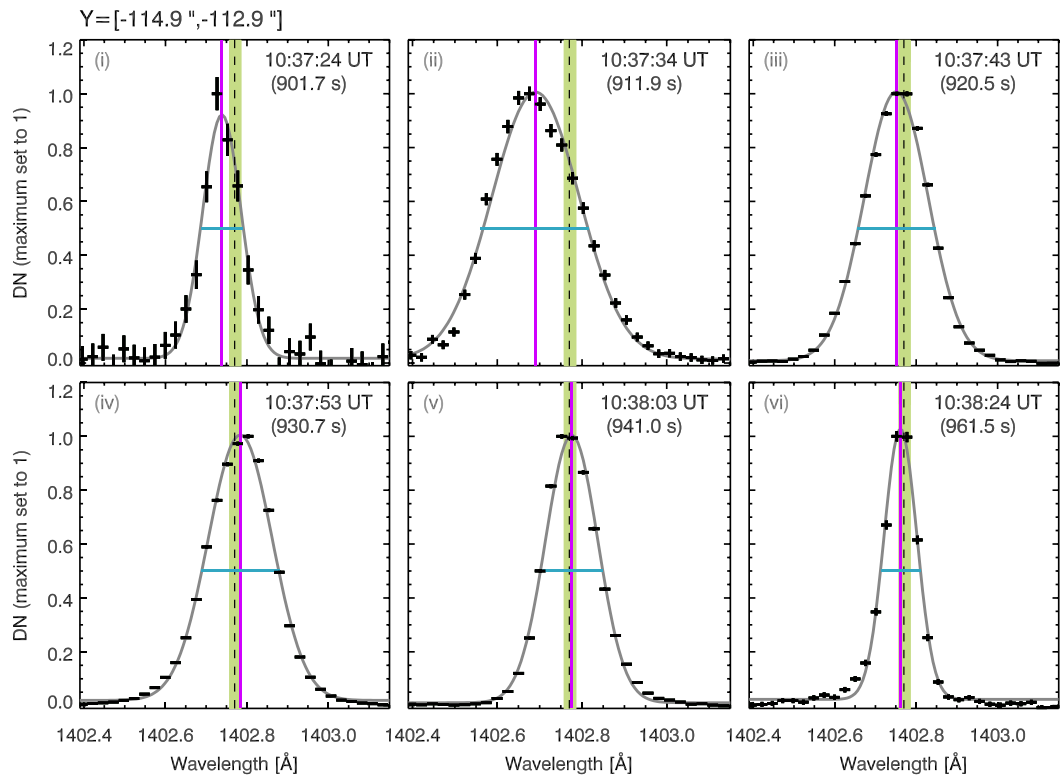

**Fig. 2.** The Si IV lines at six example flare times (black), observed over a 2" region along the north-south direction. The lines are fitted with a single-Gaussian (grey) that provides an estimate of the line centroid position (vertical purple line) and FWHM (horizontal turquoise line). We show the inferred Si IV reference wavelength plus its uncertainty (1402.775±0.020) Å; lime region, and the Si IV laboratory wavelength (1402.77 Å); black dashed line.

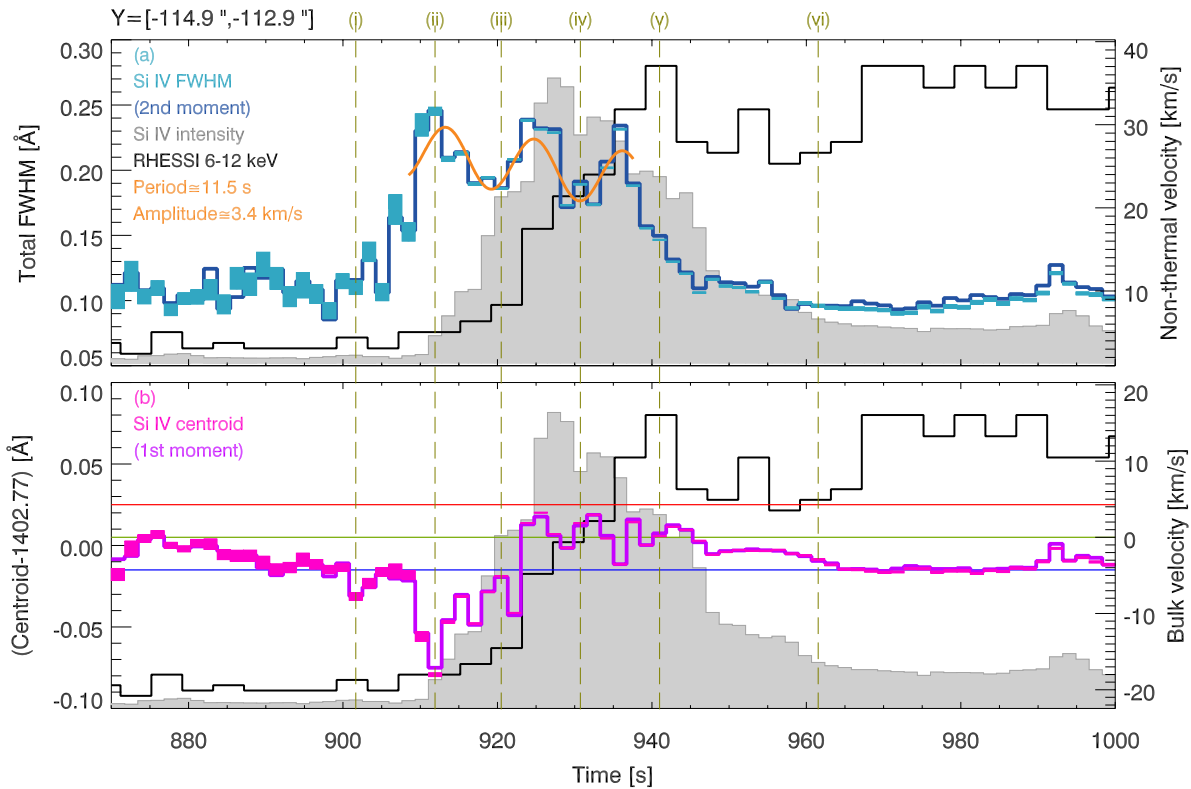

**Fig. 3.** The temporal evolution of the Si IV line properties during the flare: (a) the total FWHMs and non-thermal velocities $v_{nth}$ (turquoise), and (b) the centroid positions and bulk velocities v (purple), observed over a 2" region along the north-south direction. Both the results of Gaussian fitting and the moments analysis (see legend) give near identical results. The Si IV integrated intensity (grey), and the RHESSI 6-12 keV light curve (black), are displayed. The dashed lines indicate the times of six spectral lines shown in Fig. 2. Top panel: A sinusoidal function is fitted to the variations in $v_{nth}$, estimating the period P, and amplitude A (see legend). Bottom panel: inferred reference wavelength (lime) and uncertainty (red, blue).

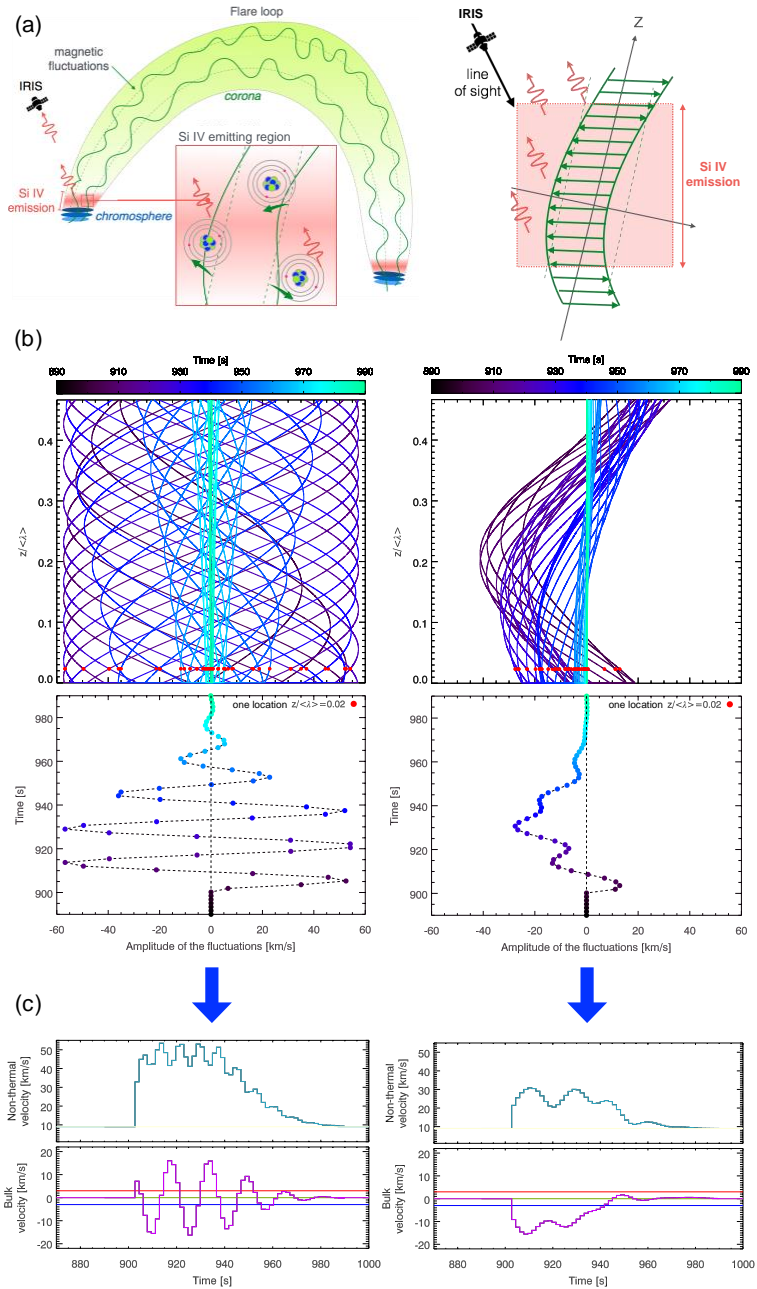

**Fig 4.** A cartoon of the flaring loop and an interpretation of the observations. (a) We envisage a situation where multiple velocity fluctuations occur in the coronal loop, *and* in the region emitting Si IV (red) before and during the flare, perpendicular to some guiding field direction z, and at some angle $>0^O$ to the line-of-sight. (b, left) We set up a wave and pass it through the modeled region, along z, representing the region emitting Si IV. As the wave passes the length of the region (denoted as $z/<\lambda>$, where $<\lambda>$ is the average wavelength of the wave), its amplitude and wavelength vary with time. (c, left) An observer (i.e. IRIS) sees all motions (velocity fluctuations) integrated along the line-of-sight, and we determine the resulting bulk velocity *(first moment)* and non-thermal velocity *(second moment)*. One wave with varying amplitude and wavelength cannot reproduce the IRIS observations; a single wave produces *clear oscillations in the bulk velocity*. (b, right) – as (b, left) but for a spectrum of 10 different interacting waves (here, representing wave interactions that can lead to turbulent dissipation). This produces a result (c, right) similar to the IRIS observation, helping us to understand the nature of plasma fluctuations in the emission region (see Materials and Methods). Movies for (b) are available (Movie 2 and Movie 3).

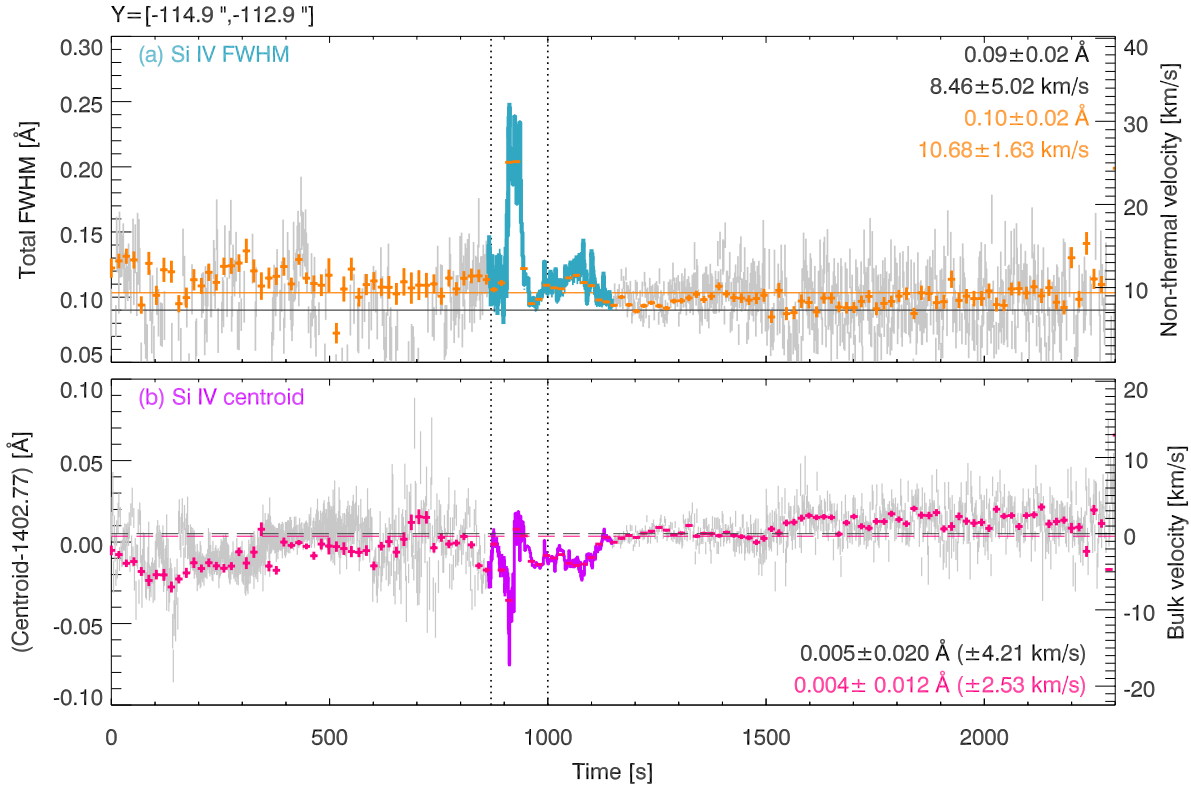

**Fig. 5.** The temporal evolution of Si IV 1402.77 Å (a) FWHM (non-thermal velocity, $v_{nth}$) and (b) centroid position (bulk velocity, v) over the spatially integrated region of Y between -114.9" and -112.9", for all times observed by IRIS during the observation (grey) and the flare time (turquoise (a) and purple (b)). In (a), the non-thermal velocity in the region before and after the flare is shown in the legend, calculated for the raster time resolution (black) and a time binned case (orange). In (b), the determined Si IV reference wavelength in the region plus its uncertainty is shown in the legend for the raster time resolution (black) and a time binned case (pink).

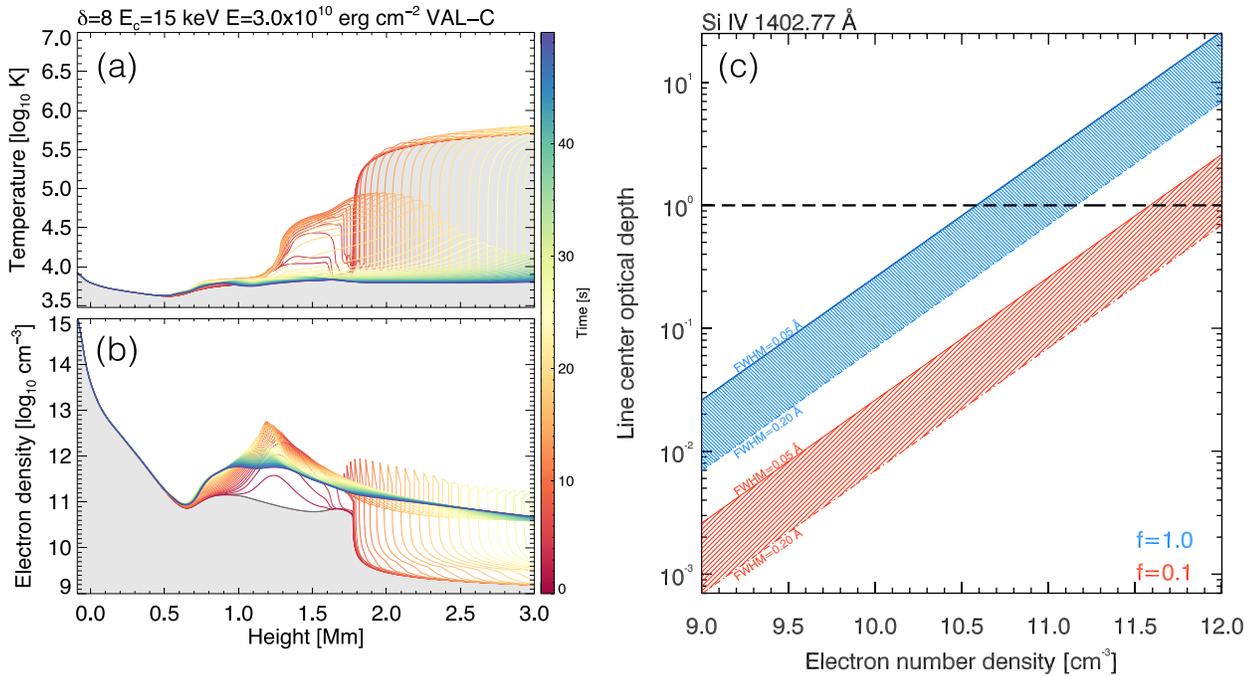

**Fig. 6.** Determining Si IV 1402.77 Å line optical depth. (a), (b) An output from the RADYN simulations showing the flare model temperature (a) and electron number density (b) versus height above the photosphere, produced using the input parameters shown. The grey region shows the quiet Sun model atmosphere (VAL-C; 30). Si IV is formed at log T=4.9 and the simulations indicate that the electron number density at this temperature ranges from log $n_e$=9.5-11.5, particularly at early times. (c): The line center optical depth, determined using $\tau_0 \sim 0.26\, f\, \langle n_e \rangle /10^{10}$ [cm$^{-3}$], versus electron number density $n_e$ for different line FWHM and filling factors f. For most cases, Si IV 1402.77 Å should be optically thin (i.e. $\tau_0 < 1$) and free of radiative absorption effects.

**Additional files:**
Movie_1.mp4 (for Figure 1)
Movie_2.mp4 and Movie_3.mp4 (for Figure 4)

**SUPPLEMENTARY MATERIALS**
Supplementary figures
Fig. S1 - A context image of the solar flare in active region 12615.
Fig. S2 - The Si IV contribution function.
Fig. S3 - A comparison of Si IV 1402.77 Å and Mg II 2796.35 Å centroid positions.
Fig. S4 - The temporal evolution of Si IV 1402.77 Å line properties during the flare (for two individual pixels).

# Supplementary Materials

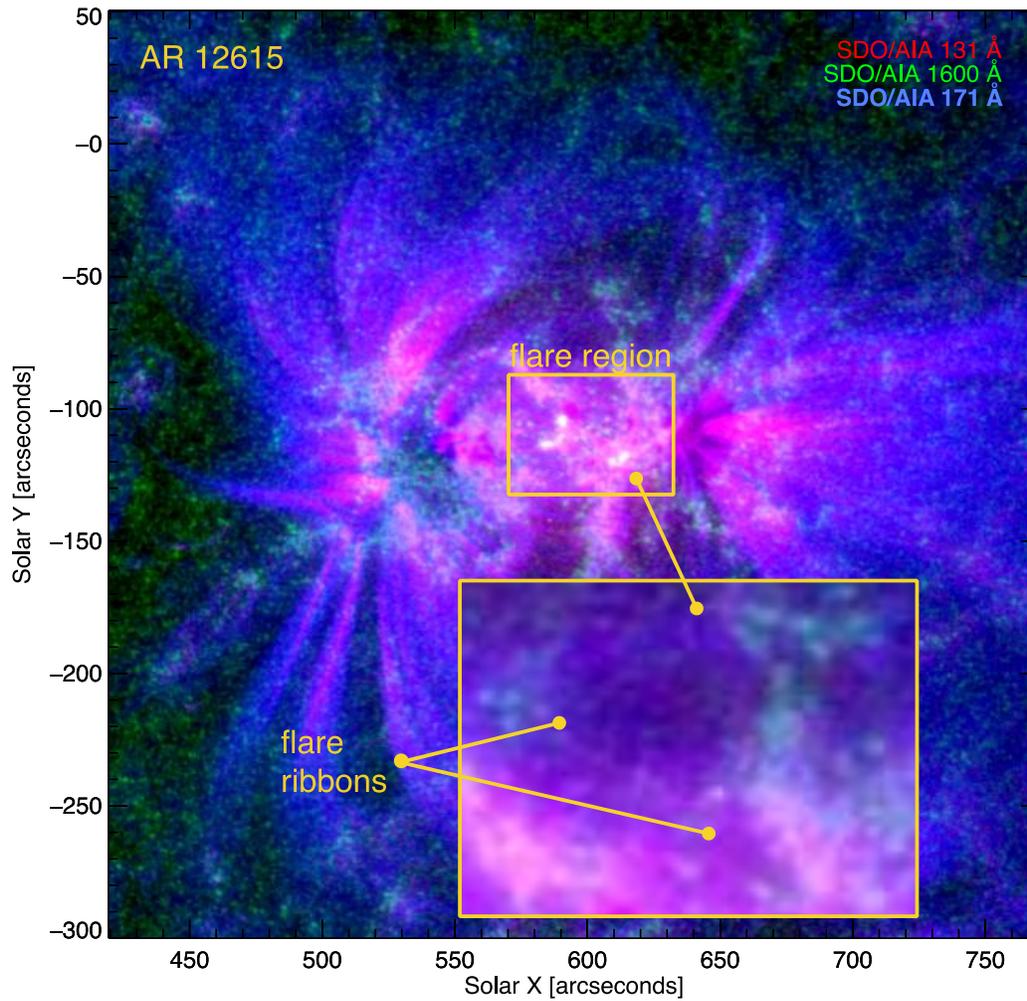

**Fig. S1.** A context image of the flare in active region 12615 using a composite Solar Dynamics Observatory (SDO) Atmospheric Imaging Assembly (AIA) image using wavelengths of 131 Å, 1600 Å and 171 Å. Zooming in on the flare region clearly shows the flare ribbons.

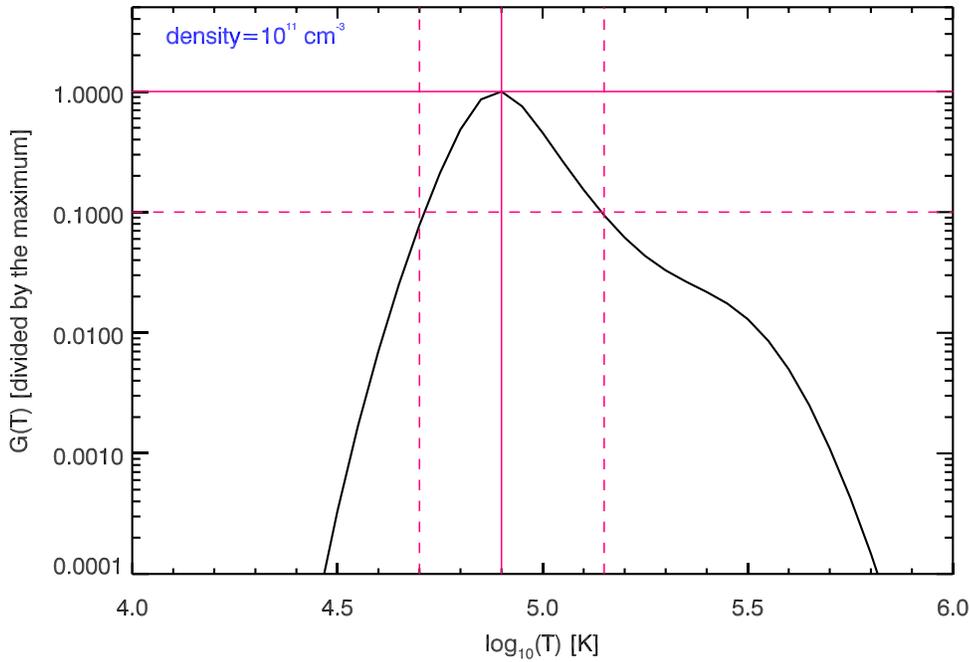

**Fig. S2.** The Si IV contribution function G(T) showing that the emission of Si IV peaks close to log T=4.9~80000 K. Over a temperature range of log T~4.7-5.15 (50000-140000) K, the emission of Si IV falls by one order of magnitude. Si IV is most likely formed close to its peak formation temperature of 80000 K, but it can be formed over a range of temperatures.

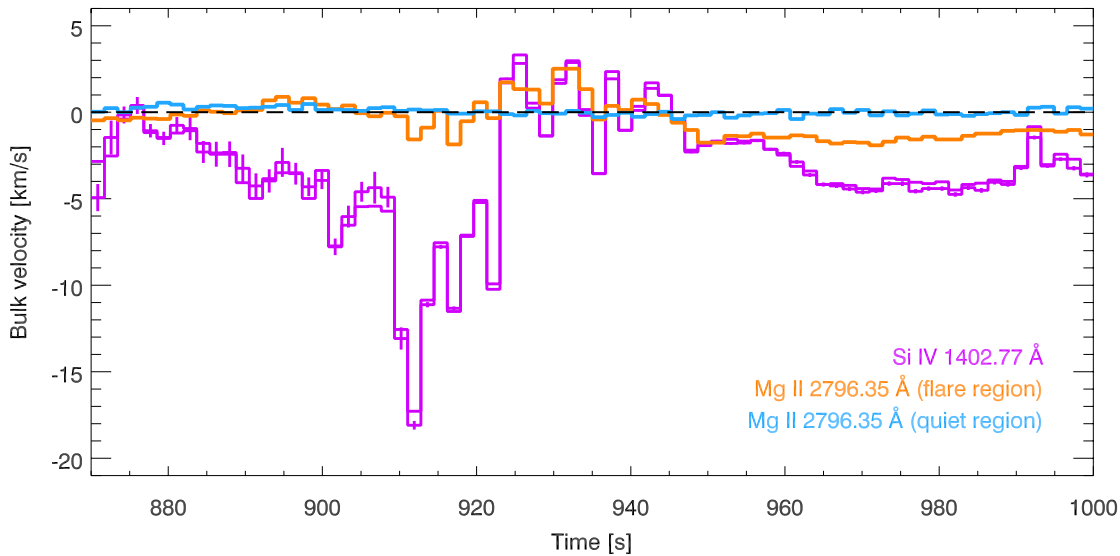

**Fig. S3.** A comparison of Si IV 1402.77 Å (purple) and Mg II 2796.35 Å (orange) centroid positions (as bulk velocities) with time. The Mg II 2796.35 Å line shows similar velocity movements to the Si IV 1402.77 Å line, but with smaller values. We can see that no sharp jumps in the centroid position are observed in the Mg II 2796.35 Å quiet region (blue).

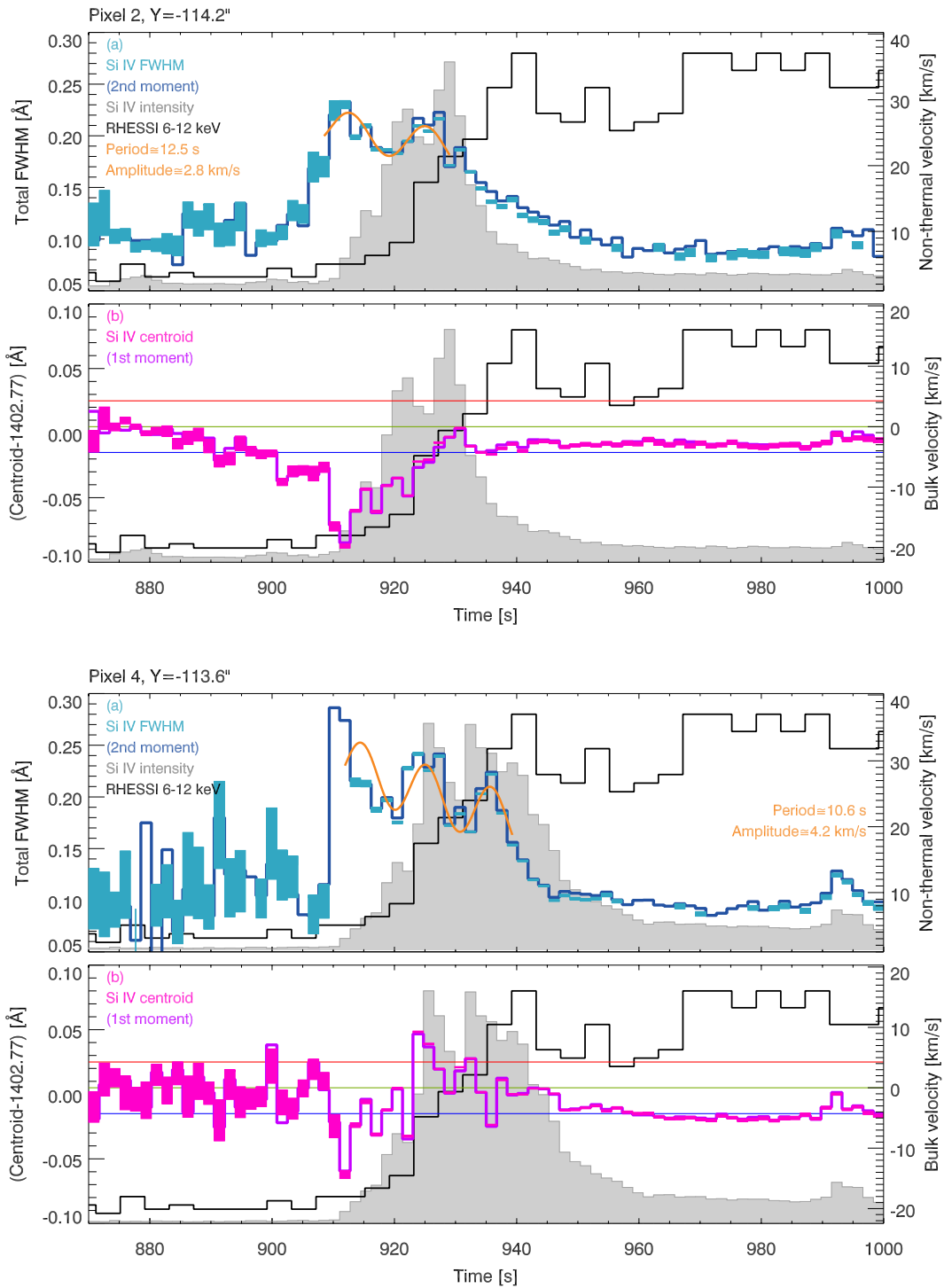

**Fig. S4.** The temporal evolution of Si IV 1402.77 Å line properties during the flare: (a) the total FWHMs and non-thermal velocities $v_{nth}$ (turquoise), and (b) the centroid positions and bulk velocities v (purple), for two individual pixels: pixel 2 at Y=-114.2" (top two panels) and pixel 4 at Y=-113.6" (bottom two panels). Both the results of Gaussian fitting and the moments analysis are shown (see legend) and both give near identical results. The Si IV integrated intensity (grey), and the RHESSI 6-12 keV light curve (black), are also displayed. A sinusoidal function is fitted to the variations in $v_{nth}$, estimating the period P, and amplitude A. The inferred reference wavelength is shown (lime line) ± uncertainty (red, blue lines).